# Numerical Solution of a Complete Formulation of Flow in a Perfusion Bone-Tissue Bioreactor Using Lattice Boltzmann Equation Method

T. J. Spencer[1], I. Halliday[1], C. M. Care[1], S. H. Cartmell[2] and L. A. Hidalgo-Bastida[2]

[1] Materials and Engineering Research Institute, Sheffield Hallam University, Sheffield, UK

[2] Dept of Chemistry, University of Manchester, Manchester, UK

**Abstract**

We report the key findings from numerical solutions of a model of transport within an established perfusion bioreactor design. The model includes a complete formulation of transport with fully coupled convection-diffusion and scaffold cell attachment. It also includes the experimentally determined internal (Poly-L-Lactic Acid (PLLA)) scaffold boundary, together with the external vessel and flow-port boundaries. Our findings, obtained using parallel lattice Boltzmann equation method, relate to (i) whole-device, steady-state flow and species distribution and (ii) the properties of the scaffold. In particular the results identify which elements of the problem may be addressed by coarse grained methods such as the Darcy approximation and those which require a more complete description. The work demonstrates that appropriate numerical modelling will make a key contribution to the design and development of large scale bioreactors.

**Introduction** Perfusion bioreactors typically comprise a squat cylindrical vessel completely occupied by a porous scaffold (see figure 1) which is a geometrically random network of PLLA struts. In an attempt to initially populate a scaffold with a critical distribution of viable, surface-adsorbed, cells and thus to form high quality implants, the tissue engineer flows a solution of cells, nutrients and waste in a controlled manner through this vessel, via carefully positioned ports. The flow is typically at a low Reynolds number, $Re = UL/\nu$, [1]. The principal determinants of bioreactor dynamics are the physical processes of transport and convection-diffusion. These mechanisms are well defined and understood [1] but the governing equations are difficult to solve for a complicated device such as a bioreactor. Currently, the maximum achievable size of bone tissue grafts is limited to a size of about 1cm diameter, and this inconvenient size limitation arises partly because of the difficulty of understanding the details of the flow processes in the device. Nevertheless, demand for more implants of better quality, larger size and of a more complex, patient-specific, shape is increasing as is the cost-effectiveness of computational resource. We believe therefore that numerical modelling of transport can make a vital contribution to future bioreactor design. In this letter, we aim to demonstrate that it is now possible quantitatively to predict species concentration, fluid velocity, wall shear stress (WSS) and species surface absorption. Notwithstanding the need carefully to cross-validate the outputs of such numerical models, our approach furnishes a multi-scale methodology which dispenses with the need for the uncontrolled approximations which hinder bioreactor modelling. Such methods are predictive and can discriminate between competing hypotheses and obtain data inaccessible to experiment. Our claim is founded on computational fluid dynamics (CFD) techniques [3] which are, at once, inherently suited to parallel computation, multi-scaling and, not least, to the handling of the complex boundary shapes as are demanded by a scaffold. Lattice

Boltzmann Equation (LBE) simulation is unlike conventional, finite element or finite volume CFD in that it solves the flow and convection-diffusion problem at the meso-scale. Put another way, in LBE correct continuum diffusion and flow physics emerge from a simple algorithm which evolves a mesoscopic distribution function (MDF) in a discretized velocity and position space [2]. Such an algorithm clearly owes much to kinetic theory [4]; indeed, it is this unique heritage which is the foundation of LBE's ability to handle multi-scale processes and geometrically complex flow or deforming boundary shapes. It is important to remark that, whilst the bioreactor flow problem is an open question (as we will demonstrate), LBE *per se* has been validated over a wide range of *Re* and of applications, against both CFD [5] and analytic calculation [6]

**Background** Mathematically, modelling transport involves solution of coupled, non-linear partial differential equations, subject to boundary conditions (BCs) which determine the surfaces on which flow and convection-diffusion must meet specified criteria [1]. The spatial extent of the boundary, the fluid considered and characteristic velocity define the scale and regime of the coupled flow problem [1]. Bioreactor modelling is inherently a multi-scale problem. Put another way, processes of vital importance to the overall efficiency and reliability of implant manufacture occur on what we define to be microscopic, scaffold-pore scales (eg. cell attachment) *and* on macroscopic scales of the device-chamber scale (eg. nutrient, cell and waste distribution). Previous work addresses only one or other of these separate scales, partitioning the modelling problem by recourse to some implicit BC which can be representative of un-modelled elements of the reactor. So, for example, macro-scale models of the whole device *average* the flow micro-scale by employing the Darcy or Darcy-Brinkman flow approximation for porous flow [7, 7a]. However, transport is indisputably described on all scales by the Navier-Stokes and coupled species convection-diffusion equations [1]. Accurate, Navier-Stokes descriptions of flow, based upon finite element method and traditional CFD, have accessed the flow mechanics of model porous media, comprised of regular, idealized voids and channels [8], and also small ($1mm^3$) sections of irregular scaffolds with fictitiously theorised continuously smoothed surfaces [8a], but neither reach the spatial scales of a whole bioreactor. Before pressing a critique, we note that such tools provide high quality data (albeit from approximate models) or illuminate effective medium theories. The work presented here can be viewed as an attempt to bound or validate these approaches. That said, for Darcy flow, $Re \ll 1$, whereas $Re = o(1)$ in extended parts of a scaffold, close to ports [9] and effective media emerge from our study as limited representations of real scaffolds. Other approaches include local, microscopic models of, for example, scaffold pore-scale motion or cell attachment using agent-based techniques [10], molecular dynamic inspired approaches [11] or cellular automata [12].

**Method** We choose to study a perfusion bone tissue bioreactor in steady flow. The system is illustrated diagrammatically in figure 1, which represents our geometrical boundary conditions, and is fully described below. Note, the plane surfaces in figure 1 represent the symmetry planes we have exploited in our computations. In order to save computational expense we have ignored the microscopic symmetry- breaking of the random scaffold. This assumption does not affect our core conclusions and is a limitation which can be simply overcome by larger scale simulations. Fluid inlet and outlet flow rates and profiles were specified only at the vessel ports, with the corresponding *Re* matched to experimental data [7, 9]. For definiteness, we chose to consider convection-diffusion of a single nutrient (oxygen), waste (carbon dioxide) and cell species, for which the appropriate diffusion coefficients were taken to be $3\times10^{-9}$, $1.6\times10^{-9}$ and $3\times10^{-13}$ m$^2$ s$^{-1}$ respectively [9,13]. Cell diameters are typically 1.5 orders of magnitude less than the diameter of a PLLA scaffold void [9] and, accordingly, we choose, in this present study, to represent cells in terms of a volume concentration, coupled to a surface concentration (equation 4) as well as to flow and other species. Accordingly, within standard notation [14], the macroscopic system of equations to describe our system of transport is the incompressible and Navier-Stokes equations for the velocity, $v$, and pressure, $p$, of a cell culture medium with shear viscosity, $\eta$, and density, $\rho$,

$$\nabla \cdot v = 0 , \qquad (1)$$

$$\frac{\partial}{\partial t} v_\alpha + (v \cdot \nabla)v_\alpha = -\frac{1}{\rho}\nabla p + \frac{\eta}{\rho}\nabla^2 v_\alpha, \qquad (2)$$

for advected / diffused, oxygen and carbon dioxide concentrations, $\phi_i$ (subscript *i* identifies species):

$$\frac{\partial}{\partial t}\phi_i + (v \cdot \nabla)\phi_i = K_i \nabla^2 \phi_i + q_i C, \qquad (3)$$

Where $K_i$ is the diffusion coefficient, $C$ the local cell volume concentration in suspension and $q_i$ are first order kinetic reaction coefficients (positive for carbon dioxide, negative for oxygen [16]), and for advected / diffused cell volume concentration,:

$$\frac{\partial C}{\partial t} + (v \cdot \nabla)C = K_c \nabla^2 C + q_c(\phi_{O_2} - \phi_{O_2}^0), \qquad (4)$$

here $q_c$ is a first order reaction coefficient and $\phi_{O_2}^0$ is the oxygen concentration at which cell volume concentration is maintained constant. The cell attachment follows a receptor-ligand model for the surface absorbed cell concentration, $C_s$,

$$\frac{\partial C_s}{\partial t} = k_a C(C_s^{sat} - C_s) - k_d C_s, \qquad (5)$$

in which $k_a$ is the association rate reaction coefficient and $k_d = \sigma'\tau$, the dissociation reaction coefficient [17].

The cell volume concentration is initially uniform and from experimental data [9] is taken to be 56 cells mm$^{-3}$. In order to recover numerically the steady-state solutions we have developed a three-dimensional, fully-coupled, whole bioreactor model without any implicit scale partition. Our flow model is based on LBE simulation [2] and uses two sets of no-slip boundary conditions, visualised in figure 1. The first, external, no-slip boundary represents the bioreactor chamber walls, flow sources and sinks, outlined above. The second, more interesting, internal, boundary derives from a micro CT scan at 16$\mu$m resolution, of an experimental PLLA scaffold. By projecting this scan over a "D3Q19" LBE simulation lattice [2], it is possible to identify the parts of the lattice which are inaccessible to fluid (see figure

1 caption). The literal and very complex internal boundary shape which results cannot be handled by any current conventional CFD technique efficiently over the entire cylindrical vessel dimensions and is in no sense an effective medium. However, simple meso-scale boundary conditions allow the LBE simulator access to the flow at a higher level of accuracy and detail than has heretofore been achieved in the modelling of bioreactors. The boundary conditions are based upon the physics of the interaction of the MDF with a solid boundary [2, 15].

In LBE, the flow problem is represented by MDFs which evolve by propagating and equilibrating on the links of a D3Q19 regular lattice system,. The MDFs interact very simply with the boundaries, however complex, to produce a time-marching solution, which was run to steady-state to produce data characteristic of both the scaffold micro-scale and whole-device macro-scales. These results are discussed in the next section. The coupled convection-diffusion problems may also be solved via LBE methods [2], *per contra*, we chose to use orthodox finite differencing methods. Additional upwinding methods are required to stabilize the low diffusion regime of the cell volume concentration in Eq. 4. To address both the pore scale and the bioreactor vessel geometry scale (*cf.* figure 1) we employ large scale parallel computing domain decomposition methods in which the bioreactor and scaffold is split into many numerically regularly sized domains, each domain calculated concurrently on separate processors.

**Results** We will here discuss the three most important aspects of our data to emerge. First, we consider the mechanical transport properties of bulk scaffold. For this study of a typical PLLA scaffold's intrinsic properties, the said scaffold was CT scanned, then cleaved *in silico*, into a regular cube, side $L$, volume 4 mm$^3$, and finally inserted into a square cross-section flow guide with an inlet and outlet both of length $L$. Pressure driven flow at $Re = 1$ (typical of bioreactors) with a Poiseuille, purely axial, inlet velocity profile impinges on the guide inlet and thence onto the cleaved surface of the scaffold. Figure 2 (a) shows only part of the inlet and outlet flow guides, note, as well as the output-normalized pressure field (below). The data of figure 2(b) shows sample velocity streaklines. Since cells, which have a density close to that of water, approximate non-inertial advected particles, this data is representative of cell trajectories in flow. The data of figure 2 (c) is an assemblage of velocity field samples, each compiled as follows. The flow velocity modulus field (expressed in lattice units) was computed and sampled over the set of three equi-spaced, parallel mathematical reference planes, all orientated perpendicular to the cleaved scaffold face and all contained within it. Figure 2(d) depicts the scaffold surface WSS distribution within the scaffold. Figure 3 shows two-dimensional cross sections from the full scale bioreactor simulations across half the central plane during the initial cell attachment. Images are representative of flow, nutrient, waste and cell distributions after 3 hours of cell and medium perfusion, flow rate 0.01 ml/min [9]. Figures 3(a) and 3(b) identify completely the flow and shear stress fields. Figures 3(c) and 3(d) show nutrient and waste respectively which are relatively unaffected by the presence of the scaffold boundary due to the higher diffusivities of the species. In figure 3(e) the cell volume concentration is shown to be much more heterogeneous and affected by the pore scale boundary conditions. Clearly the surface absorbed cell concentrations, figure 3(f), directly benefit from higher cell volume concentrations (*cf.* figure 3(e)). Cell concentrations tend to follow regions of higher velocity and show regions of low concentration in stagnation areas, due in main to the low diffusivity, and also in regions of lower (higher) nutrient (waste) concentrations.

**Discussion** The unexpected and high degree of heterogeneity present in the flow structure within the bulk scaffold is very apparent in data of figures 2 and 3. Although the PLLA scaffolds have a very porous and open structure, it is surprising that the flow spontaneously separates into well defined preferred channels (figure 2(c)) which, whilst tortuous, extend throughout the flow-wise dimension. The flow modalities predicted in figures 2 and 3 could not be predicted in any other way that by solving the Navier-Stokes and continuity equations, with appropriate BCs. These results suggest that contiguous regions of scaffold are dominated by different physical regimes of transport; some are advection dominated, others must rely upon diffusive transport of species and cells. Notably, in figure 2(d) high WSS "hotspots" are apparent, even within an ideally supplied scaffold, which hints at an irreducible degree of WSS inhomogeneity determined purely by the scaffold. In figure 3 our data on waste ($CO_2$) and nutrient (oxygen) transport are much less affected by the micro-scale boundary and would clearly be closer to Darcy behaviour. Large diffusion coefficients and, hence, small Peclet numbers associated with molecular species responsible for this behaviour of the concentration profiles, which are much smoother than flow profiles. Here, solutions compare more favourably with corresponding results obtained in Darcy regime. However, the diffusion coefficient of cells is approximately 4 orders of magnitude smaller than that for chemical species: it is for this reason that the cell transport shown in figure 3(e and f) is much more heavily influenced by those micro-scale velocity heterogeneities, identified in figure 2(b).

**Conclusion** Figures 2 and 3 of the present work demonstrate a step change in bioreactor modelling capability. For the first time, a closed, fully-coupled model of a complete bone tissue bioreactor system has produced data for experimentally relevant timescales. More specifically, these figures also demonstrate that some aspects of bioreactor dynamics, such as transport of waste and nutrient species possessed of large diffusion coefficients, and small associated Peclet numbers, can be safely predicted within the Darcy flow approximation, others, such as cell transport (described by diffusion coefficients which are 3-4 orders of magnitude smaller) clearly cannot. The latter may only be correctly understood by studying full Navier-Stokes, coupled convection-diffusion and continuity equation solutions.

Clearly, the perfusion bioreactor studied supports heterogeneous flow. Accordingly, the transport physics is diffusion or advection dominated in different regions of the scaffold, which raises obvious design issues.

The next most important aspect of the flow formulation problem appears to us to be the extent to which cells may be correctly represented as a species, especially within the context of the modelling assumptions made around the interaction of advected and surface absorbed cell species. The portable, scalable, accessible and versatile lattice Boltzmann equation deployed in this work offers further possibilities of ambitious multi-scaling. We aim next at more literal representation of cellular species as deformable, explicit, Lagrangian particles, on reactor-scale simulations.

**Acknowledgements**

The work reported here was entirely funded by BBSRC grant reference BBF0137441.

**Figures**

**Figure 1** The geometrical boundary conditions applied to out LBM computations. Boundary conditions are visualised by colour. The bioreactor vessel is black, the surface of the vessel and scaffold are yellow and the absence of colour denotes void. The scaffold (diameter 8 mm) boundary condition was constructed from CT scanned slices, projected layer by layer onto the simulation lattice. Only one quadrant of the cylindrical bioreactor vessel is shown. Fluid inlet and outlet ports (diameter 1 mm) are also shown. The plane surfaces in the figure represent the symmetry planes, surface-normal gradient on which are taken to vanish in all our computations.

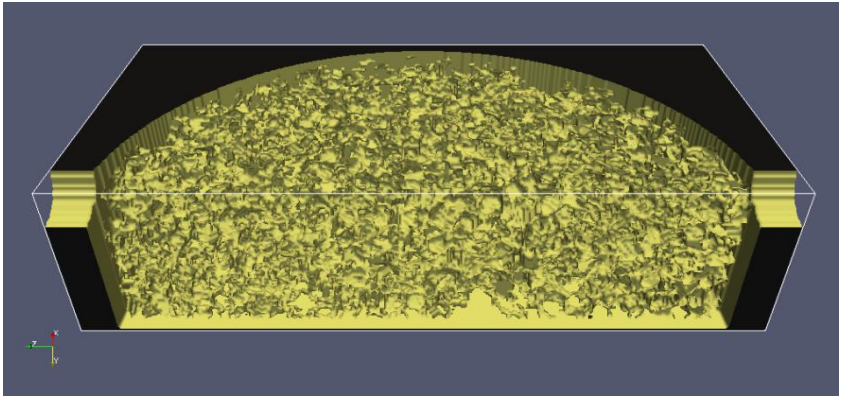

**Figure 2**

Figure 2 (a) Inlet, outlet flow guides and cleaved scaffold. The cubic sampled scaffold volume element had volume 4mm$^3$. 2(b) Velocity streaklines and relative pressure. This data is representative of cell trajectories in flow. 2 (c) An assemblage of velocity field samples in which the flow velocity modulus field (expressed in lattice units) was computed and sampled over the set of three equi-spaced reference planes orientated perpendicular to the cleaved scaffold face: streaklines of connectivity are imaged sourced at a central pore. 2(d) The scaffold surface WSS distribution within the scaffold is noticeably inhomogeneous.

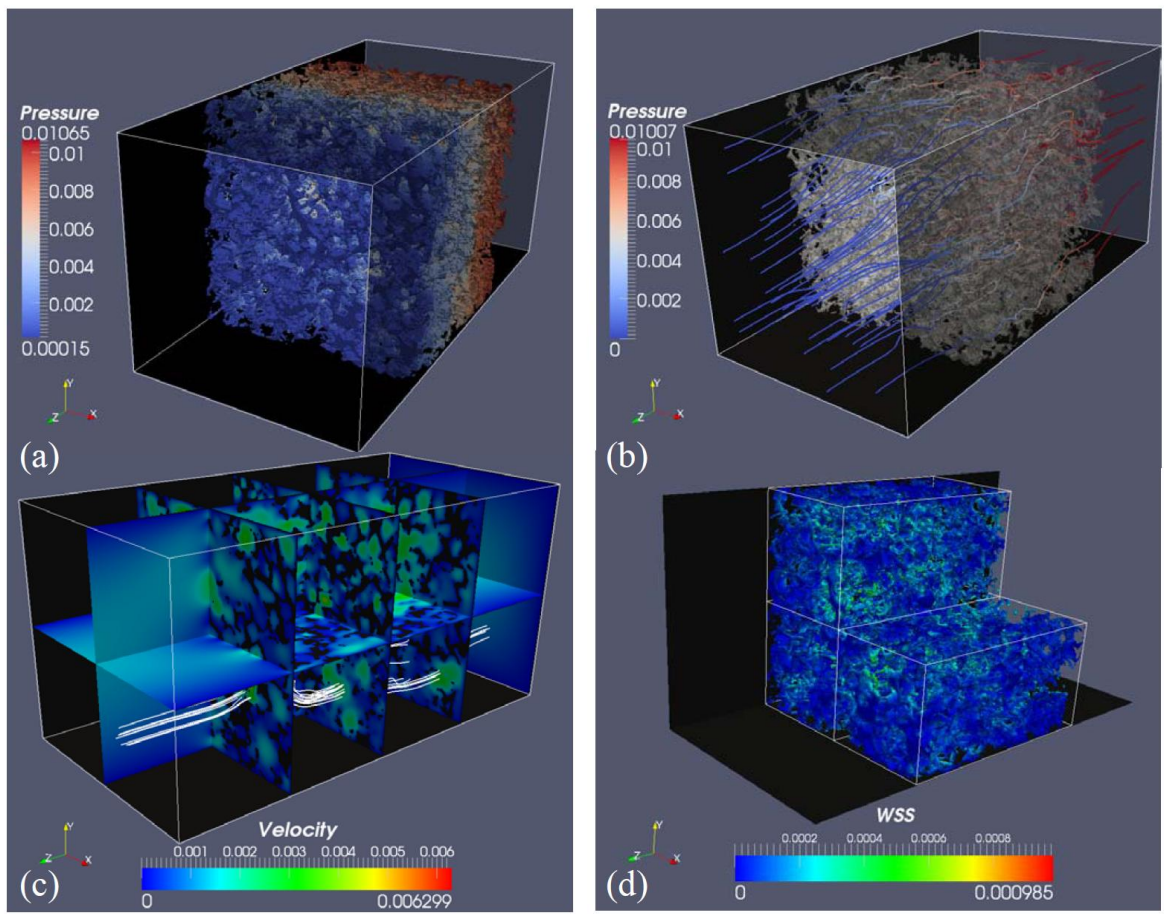

**Figure 3**

Figure 3 Two dimensional cross-section images of mass transport and cell attachment across the bioreactor from inlet (left) to outlet (right); dimensions given in figure 1. Images after a 3 hour cell medium perfusion, flow rate 0.01 ml/min. (a) Log plot of the velocity field, (b) log plot of the shear stress, (c) relatively homogeneous oxygen and (d) carbon dioxide concentrations, (e) heterogeneous advecting cell volume concentration and (f) surface absorbed cell population.

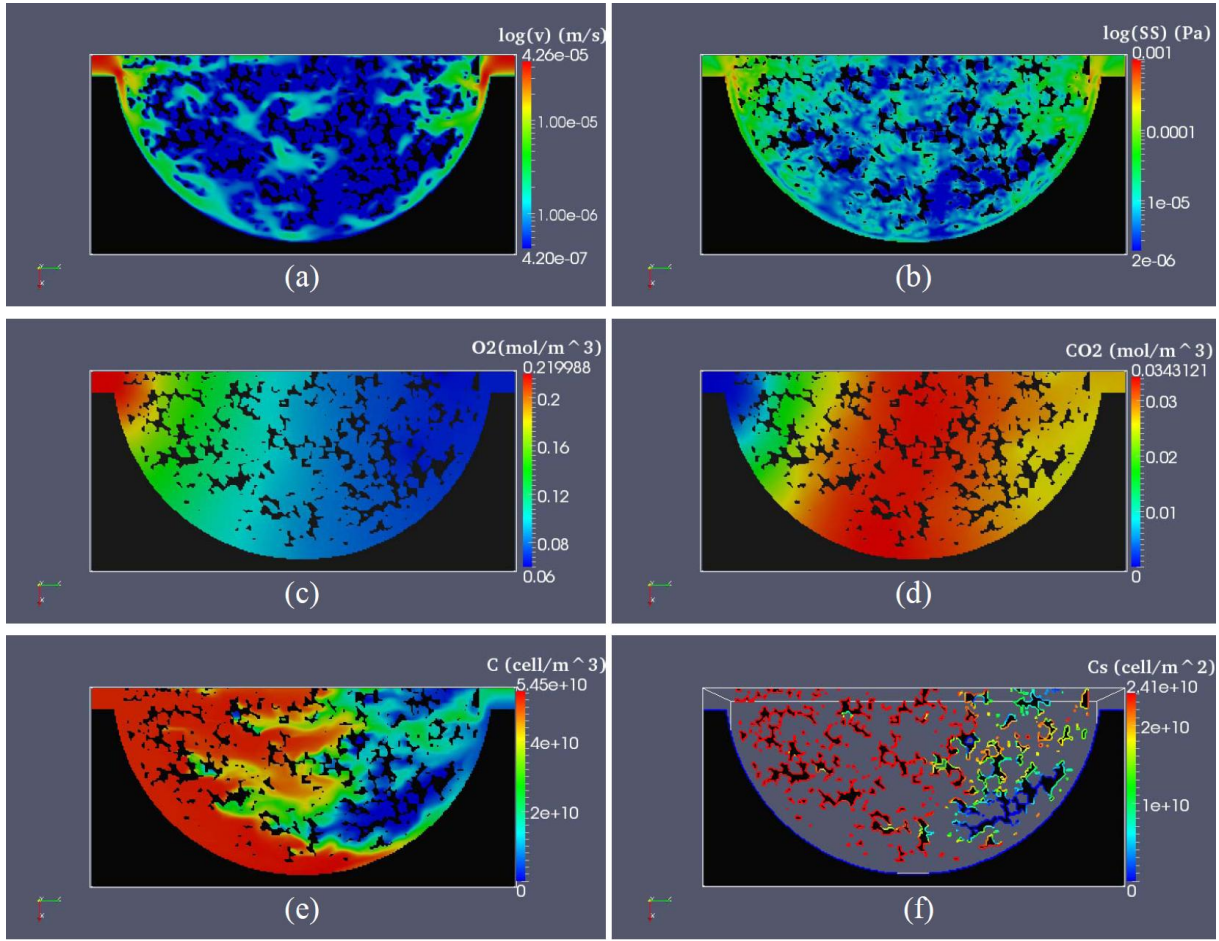